\documentclass[conference]{IEEEtran}
\usepackage{amsmath, amsfonts, amssymb}
\usepackage[amsmath]{ntheorem}
\usepackage{latexsym}
\usepackage{graphicx}
\usepackage{color}
\usepackage{subfigure}
\usepackage{balance}

\makeatletter
\renewtheoremstyle{plain}%
  {\item[\hskip\labelsep \theorem@headerfont ##1\ ##2\theorem@separator]}%
  {\item[\hskip\labelsep \theorem@headerfont ##1\ ##2, ##3\theorem@separator]}
\makeatother
\newtheorem{example}{Example}

\DeclareMathOperator*{\argmin}{arg\,min}

\begin{document}

\title{Reference Based Genome Compression}


\author{\authorblockN{B. G. Chern, I. Ochoa, A. Manolakos, A. No, K. Venkat and T. Weissman} \authorblockA{Department of Electrical  Engineering\\ Stanford University, Stanford CA 94305\\ Email: \{bgchern, iochoa, amanolak, albertno, kvenkat, tsachy\}@stanford.edu} }


\maketitle

\begin{abstract}
DNA sequencing technology has advanced to a point where storage is becoming 
the central bottleneck in the acquisition and mining of more data.  Large amounts of data are
vital for genomics research, and generic compression tools, while viable,
cannot offer the same savings as approaches tuned to inherent biological
properties. We propose an algorithm to compress a target genome given a known reference genome. The proposed algorithm 
first generates a mapping from the reference to the target genome,
and then compresses this mapping with an entropy coder. As an illustration of the performance: applying our algorithm to James Watson's genome with {\em hg18} as a
reference, we are able to reduce the 2991 megabyte (MB) genome down to 6.99 MB, while Gzip compresses it to 834.8 MB.\\

\end{abstract}

\IEEEpeerreviewmaketitle

\section{Introduction} \label{sec:intr}
Advances in genomics over the past decade have significantly reduced the amount
of time and money required to sequence individual genomes.  What used to take years
and billions of dollars can now be done in a matter of days for only thousands
of dollars.  As a result, the amount of genomics data has expanded to the point
where data storage is now the bottleneck in the sequencing process.  While
general purpose compression algorithms such as gzip can provide good
compression, much more can be achieved by taking into consideration the
inherent properties of DNA.  For instance, while the human genome consists of
roughly 3 billion base pairs, it is well documented that any two human genomes
are more than 99\% identical \cite{nature}.  At the same time, there is an imminent need for efficient compression and storage of the increasing amount of genomic data. The similarity in genomic data suggests compression schemes that exploit this
redundancy.  Starting from this premise, we investigate the problem of lossless
compression of a target genome when a reference genome is available to both the
encoder and decoder without the aid of any additional external information.

Lossless genome compression with a reference sequence can be viewed as a two
stage problem. In the first stage, a mapping from the reference
genome to the target genome is generated in an efficient manner. The second stage then
involves describing this mapping concisely to the decoder.  Given the mapping
and the reference genome, the decoder is thus able to recover the target genome
sequence.

The process of constructing a mapping from the reference genome to the target
genome in a concise manner can be quite challenging. In \cite{wheeler} 
this problem is approached from a biological standpoint. Given a reference and target
genome, two separate files are produced: one consisting of single nucleotide
polymorphisms (SNPs), and one consisting of insertions and deletions of
multiple nucleotides (indels). Thus, the mapping between genomes is expressed
as a combination of insertion, deletion and substitution operations. Working
with these two files and an additional database of common SNPs for the human
genome, \cite{dnazip} presents an algorithm to losslessly compress the SNP and indel files.

In practice, the SNPs and indels files may not be available and generating them is an overhead, especially when it does not
necessarily determine the smallest set of differences between the genomes. This
drawback motivates us to develop an end-to-end compression scheme that seeks to generate
an efficient mapping between a reference and a target genome, and then describe this mapping in an efficient manner.

This problem and its variations have been studied in \cite{green}-\cite{tse}. In \cite{green} Pinho et al. present GReEn, a compression tool based on arithmetic coding that overcomes some drawbacks of the previously proposed tool GRS \cite{grs}. In \cite{rlz1} Kuruppu et al. introduce RLZ, an algorithm that uses a greedy technique based on LZ77 \cite{lz} to parse the target genome into factors, each of which occurs in the reference sequence. An optimized version of this algorithm is proposed in \cite{rlz2}. In \cite{manipulation} Heath et al. focus not only on compression, but also on efficient manipulation, studying the tradeoff between the two. In \cite{tse} Ma et al. propose an algorithm to compress a source sequence (target genome) given side information (reference genome) and show that it is asymptotically optimal for small insertion and deletion probabilities, assuming i.i.d. deletions. 

Typically, the transitions between a reference and a target genome can be described by a combination of insertion, deletion and substitution operations. Due to the nature of this corruption mechanism, the assumption of the data sequences being jointly stationary is not valid\footnote{To see why joint stationarity is not a valid assumption, it suffices to note that even for the simple scenario where the reference sequence is stationary and the target sequence is its corrupted version under i.i.d. deletions, the process pair need not be jointly stationary.} and, as a result, existing schemes for compression in the presence of side information, which are designed under that assumption, are not applicable. For example, the LZ77 algorithm \cite{lz} and the Context-Tree Weighting (CTW) method \cite{ctw}, along with their extensions to side information \cite{lz2} and \cite{ctw2}, respectively, are well studied compression techniques. The optimality of \cite{lz2} and \cite{ctw2} is under the jointly stationary model. Further, DNA data has a much higher proportion of substitutions, than insertions or deletions. For instance, in the SNP and indel files presented in \cite{dnazip}, it can be seen that substitutions constitute more than $90\%$ of the operations.

Motivated by the LZ77 algorithm, and under the premise of a large fraction of substitutions, the first stage of our proposed algorithm parses the target genome using the reference as side information. Subsequently, the proposed algorithm segments the differences into substitutions, and certain types of insertion and deletion operations - a categorization which assists the second stage of the algorithm where these differences are compressed. 

The rest of this paper is organized as follows:  In Section \ref{sec:problem} we formally describe the problem. In Section \ref{sec:alg} we describe the proposed algorithm, and in Section \ref{sec:results} we discuss its performance. Section \ref{sec:conclusion} concludes the paper.

\section{Problem Description}\label{sec:problem}

We begin by introducing the notation. Let upper case, lower case, and calligraphic letters denote,
respectively, random variables, specific or deterministic values they may assume, and their alphabets. $X_m^n$ is shorthand for the $n-m+1$ tuple $\{ X_m, X_{m+1}, \ldots, X_{n-1}, X_n\}$. $X^n$ also denotes $X_1^n$. When $i \leq 0$, $X^i$ denotes the null string as it is also for $X_i^j$, when $i > j$. $X^{n \backslash i}$ denotes  $\{ X_1, \ldots, X_{i-1}, X_{i+1}, \ldots, X_n\}$. The cardinality of an alphabet $\mathcal{X}$ is denoted by $|\mathcal{X}|$.  

We consider the problem of compressing the target genome $X^N$ given the reference genome $Y^M$ as side information, which is available at both the encoder and decoder. We have\footnote{Typically, genomic data comprises of the characters $\{A,C,T,G,N\}$, though it may vary from case to case.} $\mathcal{X} = \mathcal{Y}$. The encoder is described by the mapping $f(X^N, Y^M)$ which indicates the compressed version of $X^N$ given $Y^M$ as side information. The decoder is described by the mapping $g( f(X^N, Y^M), Y^M)$. Since our framework is that of lossless compression, the encoder-decoder pair satisfies
\begin{equation*}
X^N = g( f(X^N, Y^M), Y^M).
\end{equation*}

The objective is to construct an encoding/decoding scheme that minimizes the length of the representation $f(\cdot, \cdot)$,  while ensuring a perfect reconstruction of the sequence $X^N$ at the decoder.

\section{Proposed Algorithm}\label{sec:alg}
The encoding algorithm can be divided into two stages. First we generate a mapping from the reference genome $Y^M$ to the target genome ${X}^N$. In the second stage, this mapping is compressed losslessly using an entropy coder. The decoder then takes the compressed representation of this mapping and decompresses it, obtaining the mapping from ${Y}^M$ to ${X}^N$. Since ${Y}^M$ is available at the decoder as side information, it can recover ${X}^N$ perfectly by inverting the original mapping. 

We now describe the two stages performed by the algorithm.

\subsection{Generating the mapping}
The mapping generation is motivated by the sliding window Lempel-Ziv algorithm LZ77 \cite{lz}. In LZ77, new data is expressed in terms of previously parsed data. In order to exploit the similarities between the reference and the target sequences, we express strings in the target sequence in terms of a corresponding matching string in the reference sequence. As motivated in Section \ref{sec:intr}, we apply techniques which exploit the appearance of a larger fraction of substitutions. This, coupled with the fact that the reference and target genomes are not jointly stationary, allows us to get a significant improvement over \cite{rlz1}, where an LZ77 based parsing scheme is also considered. We begin by describing the dynamic window based string matching algorithm. This is followed by a segmentation of the results obtained, into substitutions, and certain forms of insertions and deletions, which facilitates the compression of the data. 

Define $y_{W_k- L_k}^{W_k + R_k }$ as the window at time $k$. Later we will discuss how to choose the parameters $(W_k,L_k, R_k)$. The basic idea of the algorithm can be summarized as follows: Assume $x_1^{n_{k}}$ has already been encoded, for some $n_{k}$. Then, find the largest length $l_k$ s.t. $x_{n_{k} + 1}^{n_{k} + l_k} =y_{i}^{i + l_k - 1}$ and $x_{n_k + l_k + 1} \neq y_{i + l_k }$, for some $W_k-L_k \leq i \leq W_k+R_k$, and encode $i, l_k$ and the novel symbol $x_{n_k + l_k + 1}$. Now, set $n_{k+1} = n_k+l_k+1$ and repeat the procedure. Thus, starting from a position in the target sequence, we find the longest matching string within a fixed window in the reference sequence. We then encode the position and length of the match in the reference as well as the first unmatched novel symbol. At the end of this part of the algorithm we have a set of instructions $ \{ F \} $ that suffices to reconstruct the target genome based on the reference.

Ideally, we would like to allow the encoder to search for a match in the entire reference sequence $Y^M$ in order to find the best match. However, for computational tractability we restrict our algorithm to a finite window search, while dynamically updating its position and width along the reference so as to efficiently capture the synchronization in traversing along the two sequences. This process is described in detail below. 

The window at time $k$ is described by the parameters $(W_k,L_k,R_k)$. Intuitively, the center of the window $W_k$ is the rough estimate of where we expect to find a match in the reference sequence, while accounting for shifts in synchronization due to insertions and deletions. We choose $(W_{n_k},L_{n_k},R_{n_k})$ as a function of the previous $(W_{n_i},L_{n_i},R_{n_i})$, $l_i$ and $p_i$ (where $i<k$). Note that we are not defining window parameters for all $n$, but only for specific time instant $n_k$ recursively. In practice, we use this idea to find optimal $W_{n_k}$'s and use fixed $(L_k,R_k)$.

The main idea is the following: First, store the previous $M$, $l_i$ and
$p_i$, and increase $W_{n_k}$ by $l_k$. Then, if $p_i$ increases (or decreases)
too much, shift $W_{n_k}$ so as to keep $p_i$ near the window center to keep
the window size small. More precisely, check the previous $M$ matches and shift
$W_{n_k}$ by the median of these $p_i$. Here we take the median of $p_i$ since
we want to be conservative and avoid excessive shifts, and to allow the
algorithm to be agnostic to bursty insertions and deletions.
In order to keep this algorithm practical, we do this correction every $M$
matches. This is a trade-off between a reasonable running time and improved window synchronization. Note that this idea is equivalent to estimating the deletion and
insertion rates at each update-time based on a certain number of previous matches. 

Define the set $S(n,W,L,R,x^\infty, y^\infty) = \{l :  x_{n + 1}^{n + l} =y_{i}^{i + l - 1} \mbox{and } x_{n + l + 1} \neq y_{i + l}, \mbox{for some } W - L \leq i \leq W + R\}$. Then, the procedure can be formally described as follows:

\begin{tabbing}
\rule{\linewidth}{0.4pt}\\
\textbf{Algorithm to generate the mapping}
\\*[\smallskipamount]
\textbf{Initialize:} Set $n_1 =0$ and $k = 1$
\\*[\smallskipamount]
\textbf{Do:} \\
\quad 
\= 1. Compute $S_k = S (n_k, W_k, L_k, R_k, x^\infty, y^\infty)$\\
\> 2. Denote the k-th phrase by $x_{n_{k} + 1}^{n_{k} + l_k+1}$, where $l_k = \max S_k$\\
\> 3. Set $p_k = \argmin_i S_k$ and $z_k = x_{n_k + l_k + 1}$\\
\> 4. Store $F_k =  (p_k, l_k, z_k)$ \\
\> 5. Set $n_{k+1} = n_k + l_k + 1$,  $k=k+1$ and update the \\
\> window parameters $(W_k, L_k, R_k)$\\
\> 6. Repeat the process until the sequence $x_1^N$ is exhausted\\
\rule{\linewidth}{0.4pt}
\end{tabbing}

\begin{example}\label{ex:mapping}
Consider the following example of a target genome and a reference genome.
\begin{align*}
\mbox{Target : }&\underline{AATG}C\underline{AGGTAC}T\underline{ATAAG}N\underline{AA}N\underline{TGC\cdots}\\
\mbox{Reference : }&\underline{AATG}T\underline{AGGTAC}\underline{AT\underline{AA}G}A\underline{TGCNNNN\cdots}
\end{align*}
In this case, the set of instructions $\{F\}$ will be as follows,
\begin{align*}
F_1 : (p_1,l_1,z_1) =& (1,4,C)\\
F_2 : (p_2,l_2,z_2) =& (6,6,T)\\
F_3 : (p_3,l_3,z_3) =& (12,5,N)\\
F_4 : (p_4,l_4,z_4) =& (14,2,N)\\
\vdots&
\end{align*}
\end{example}

Notice that the set of instructions $\{ F \}$ given the reference genome as side information suffices to perfectly reconstruct the target genome. We perform a further step to explicitly identify insertions, deletions and substitutions that represent the edits from the reference to the target genome.  Working under the premise that most of the differences between two genomes are given by substitutions, this step is designed towards compression of such genomes by reducing the number of elements to store.

In other words, the aim is to reduce the size of $\{ F \} $ by merging some of the instructions and creating new ones in the form of substitutions, deletions and insertions, that are stored in the sets $\{S\}, \{I\}$ and $\{D\}$, respectively. We seek only to obtain insertions of length one, so that for the substitutions and insertions we only need to store the position at which they occur in the target genome and the new symbol. For the deletions, we need to store the position as in the previous cases, and its length, which can not exceed a predetermined maximum value $L_{max}$. Next, we explain how to classify the edits into these categories.

In order to identify the substitutions we check if the following condition between $F_k$ and $F_{k+1}$ is satisfied:
\begin{equation}
p_k + l_k +1 = p_{k+1}. \label{eq:subs}
\end{equation}
\noindent If (\ref{eq:subs}) holds, we have
\begin{equation}
x_{n_k + 1}^{{n_{k+1} + l_{k+1}}\backslash{n_{k+1}} } = y_{p_k}^{{p_{k+1} + l_{k+1} -1}\backslash{p_{k+1} -1}},
 \end{equation}
 and 
 \begin{equation}
 x_{n_{k + 1}} \neq y_{p_{k +1} - 1},
 \end{equation}
 \noindent which represents a substitution at $x_{n_{k+1}}$ from $y_{p_{k +1} - 1}$ to $z_k$. Then, we substitute instructions $F_k$ and $F_{k+1}$ by a unique instruction given by $(p_k, l_k + l_{k+1} + 1, z_{k+1})$, and we add the new substitution to the set $\{S\}$ as 
 \[ (p^{(s)}, z^{(s)}) = (n_{k+1}, z_k), 
 \]
 \noindent where $p^{(s)}$ and $z^{(s)}$ indicate the absolute position in the target genome and the new character, respectively.
  \noindent  Insertions of length one occur if the following condition holds:
\begin{equation}
p_k + l_k = p_{k+1}.
\end{equation}\label{eq:ins}
\noindent If this is the case, we have 
\begin{equation}
 x_{n_k}^{n_{k+1} + l_{k+1}} = y_{p_k}^{p_{k+1} + l_{k+1} -1},
 \end{equation}
 and an insertion of symbol $z_k$ at position $x_{n_{k+1}}$. As before, if this condition holds, we substitute the two original instructions by $(p_k, l_k + l_{k+1}, z_{k+1})$, and add the new insertion 
 \[
 (p^{(i)}, z^{(i)}) = (n_{k+1} - 1, z_k)
 \]
 \noindent to the set $\{I\}$. 

\noindent Finally, the deletions are found by checking if
\begin{equation}
2 \leq p_{k+1 } - (p_k + l_k + 1) \leq L_{max},    
\end{equation}
 \mbox{ and }
\begin{equation}
z_k =  y_{p_{k+1} - 1} 
 \end{equation}
are satisfied, meaning that we can reconstruct $x_{n_k + 1}^{n_{k+1} + l_{k+1}}$ from $y_{p_k}^{p_{k+1}+l_{k+1}-1}$ by deleting $y_{p_k + l_k}^{p_{k+1}-2}$. Therefore, the two instructions become $(p_k, p_{k+1} + l_{k+1} - p_k, z_{k+1})$, and we add the deletion 
\[ (p^{(d)}, l^{(d)}) = (n_k + l_k, p_{k+1}  - 1 - p_k - l_k)\]
\noindent  to the set $\{D\}$, where $p^{(d)}$ and $l^{(d)}$ represent the position in the target genome at which the deletion occurs and its length, respectively.

The method described above is applied to any number of consecutive instructions that satisfy a specific condition. It is straightforward to see how the new instructions are generated in that case, and we omit the description here.

Considering the Example \ref{ex:mapping} again, we can update the instructions as follows.
\addtocounter{example}{-1}
\begin{example}[continued]
\begin{align*}
\mbox{Target : }&\underline{AATG}C\underline{AGGTAC}T\underline{ATAAG}N\underline{AA}N\underline{TGC\cdots}\\
\mbox{Reference : }&\underline{AATG}T\underline{AGGTAC}\underline{AT\underline{AA}G}A\underline{TGCNNNN\cdots}
\end{align*}
In this case, the new set of instructions $\{F\},\{S\},\{D\},\{I\}$ is as follows,
\begin{align*}
F_1 :& (p_1,l_1,z_1) = (1,16,N)\\
S_1 :& (p^{(s)}_1,z^{(s)}_1) = (5,C)\\
I_1 : & (p^{(i)}_1,z^{(i)}_1) = (11,T)\\
F_2 :& (p_2,l_2,z_2) = (14,2,N)\\
\vdots&
\end{align*}
\end{example}

The decoder, with the sets $\{F\}$, $\{S\}$, $\{I\}$ and $\{D\}$ and the help of $y^M$ can recover $x_1^N$. \\

\subsection{Entropy coder}

In this subsection we describe the entropy coder that we employ to compress and efficiently describe the sets $\{F\}$, $\{S \}$, $\{I\}$ and $\{\ D \}$. Specifically, we need to store all the characters and integers that appear in those sets, each of which we treat separately. 

Recall that each instruction in the set $\{F\}$ has two integers, to represent the position and the length, respectively. For the integers representing the position we perform delta encoding, i.e., for each position we encode the difference between that position and the previous one. In addition, since they may not appear in increasing order, we keep one bit for each integer to specify the sign. For the lengths we do not perform any delta encoding. On the other hand, the integers in the sets $\{S,I\}$ appear in increasing order and thus we perform delta encoding without retaining the sign bit. Finally, for each entry of the set $\{D\}$ there are two integers, which respectively describe the position and the length of each deletion. Note that the first list of integers is ordered and therefore we perform delta encoding. We then add the deletion lengths to the resulting list. Using this technique, we have now created one list of all the integers that we need to describe to the decoder. Finally, we compress this list using Huffman encoding \cite{huffman}.

In order to address the issue of a large codebook in the associated Huffman code, we employ the following measures which we found to work well on actual genomic data, as demonstrated below.  First, it is likely that the codebook will have a sequence of consecutive integers $1,2, \ldots, N$, where $N$ can be quite large. In such a case,  we only need to store the codewords of these consecutive numbers without sending any bits to describe the actual symbols (integers), except the number $N$. The remaining integers bigger than $N$ are sorted in increasing order. Then, we perform delta encoding on them to reduce the number of integers that we need to actually describe. Finally, we perform Golomb encoding \cite{golomb} because the occurrence of small integer values in the codebook is significantly more likely than large values, especially after performing delta encoding.

As for the characters, we use fixed Huffman tables. Usually only the characters $A,C,T,G,N$ appear in the sequences, and if any other character appears, it does with very low frequency. Therefore, we use shorter bit strings for these more common characters.

\section{Compression Performance}\label{sec:results}

To experiment with the performance of our algorithm, we employed it on real genomic data. Specifically, we use the \emph{hg18} release from the UCSC Genome Browser, the korean genomes \emph{KOREF20090131} and \emph{KOREF20090224} \cite{korean}, the genome of a Han Chinese referred to as \emph{YH} \cite{chinese} and the Watson \emph{JW} genome \cite{wheeler}. The latter is generated from the edits provided in \cite{dnazip} using \emph{hg18} as the reference genome. We also apply our algorithm to two versions of the genome of the thale cress Arabidopsis thaliana, \emph{TAIR8} and \emph{TAIR9} \cite{tair}, and of the genome of the rice Oryza sativa \emph{TIGR5.0} and \emph{TIGR6.0} \cite{tigr}.

For ease of notation, hereafter we will refer to the genomes  \emph{KOREF20090131} and \emph{KOREF20090224} by \emph{KO131} and  \emph{KO224}, respectively. 

Some of the genomes present the characters in upper and lower case. Notice that our algorithm treats both of them as the same character, i.e., 'a' and 'A' are considered equivalent. For fair comparison with other algorithms, we first convert all the above mentioned genomes to upper case. 

In \cite{green}, the authors compare the performance of their proposed algorithm GReEn with RLZ \cite{rlz1} and GRS \cite{grs}, demonstrating that it outperforms the latter two in most cases. Furthermore, RLZ only handles the characters \{A, C, T, G, N\} and GRS is unable to compress sequences that differ significantly from the reference. For these reasons, we restrict the comparison of the proposed algorithm to that of GReEn.

\begin{table}
\begin{center}
\begin{tabular}{| c | c | c  | c  | c |c |}
  \hline
 Reference & Target & Size of & Gzip & GReEn & Proposed \\
  genome & genome & Target  &  &   &  Algorithm  \\

  \hline
  hg18 & JW & 2,991  & 834.8 & 18.23 & \textbf{6.99} \\
  hg18 & YH  & 2,987  & 832.0 & 9.85 & \textbf{8.50} \\
  KO224 & YH & 2,987  & 832.0 & 10.06 & \textbf{9.54} \\
  KO131&YH& 2,987 & 832.0 & 10.27 & \textbf{9.78} \\
  KO131&KO224 & 2,938 &  831.4 & \textbf{1.31} & 1.53 \\
  KO131&hg18 & 2,996  & 836.2 & \textbf{7.82} & 9.19 \\
  KO131&JW & 2,991 &  834.8 & 23.23 &  \textbf{10.22}\\
  KO224&JW & 2,991 & 834.8 & 22.97 & \textbf{10.37}\\
\hline
TAIR8 & TAIR9 & 113.63  & 34.1 & 0.0063 &  \textbf{0.0034} \\
TIGR5.0 & TIGR6.0 & 355.07 & 108.67 & \textbf{0.12} & 29.25 \\

  \hline
\end{tabular} 
\caption{Compression results in MB of the target genome for Gzip, GReEn and the proposed algorithm. Best result is shown in bold.}
\label{tab:results} 
\end{center}
\end{table}

Table~\ref{tab:results} summarizes the results of running our algorithm and GReEn for different choices of reference and target genomes among those described above. We set the parameters $M$ and $L_{max}$ to $100$ and $1000$, respectively. We also include the results obtained by the general compression tool Gzip to show the gains obtained by compression algorithms designed specifically for the problem under consideration. All results were obtained using an Intel Core i7 CPU @ 2.80 GHz with 3087 MB RAM and Ubuntu 10.04 LTS.



As can be seen, our algorithm outperforms GReEn in seven out of the ten cases where we tested both algorithms. Specifically, it is worth noticing the compression result of \emph{JW} given \emph{hg18} as the reference, in which our scheme outperforms GReEn by more than $60\%$. As a comparison, in \cite{dnazip}, the authors compress the edits of the same data set and they manage to decrease the size of the compressed file down to $7.51$ MB. Note that in our case, we do not rely on that information and we use only the initial sequence of genomes. Still, the proposed approach manages to reduce the target genome down to $6.98$ MB. Furthermore,  our algorithm performs especially well with the data \emph{TAIR}; it reduces the size  of the compressed files by $50\%$ more than GReEn.


On the other hand, notice that if the synchronization between the target and the reference genome is lost due to a significant proportion of bursty insertions and deletions, the number of generated instructions is significantly greater, leading to a large compressed file. This is the case for the data \emph{TIGR}.

Our study thus far has focused on the compression performance of our proposed scheme. Another criterion of obvious importance is the complexity. Current running times of our scheme are comparable though usually longer than those of GReEn. However, our current implementation is far from optimized towards reduction of the running time. 
Our work in progress  is dedicated to realizing what should be significant potential for improvement of the running time due, among other factors, to the highly parallelizable nature of the algorithm.

\balance

\section{Conclusion - Future Work}\label{sec:conclusion}

In this paper we investigate the problem of compressing a target genome sequence given a reference genome that is known at both the encoder and decoder. We exploit the redundancy between two genomes to effectively compress the differences between the target and reference sequences. We present an efficient compression scheme which does not rely on any other external information. The proposed algorithm, which is motivated by the sliding window Lempel-Ziv algorithm, first generates a mapping from the reference to the target genome, and then compresses this mapping. The algorithm is end-to-end, and has the additional benefit of identifying SNP's (substitutions) which are of significant biological interest. As an illustration of our results, we compress the \emph{Watson} genome from 2,991 MB to 6.99 MB using the \emph{hg18} release as the reference.

We envisage our algorithm as having an impact on personal genomics and medicine, where storage and accessibility of patients' genome is a concern given the large sizes of human genomic data. 

Directions for future work include optimizing the choice of window parameters, to avoid losing synchronization between the target and the reference genome when the rate of the deletions and insertions is high. Other considerations include reducing running time, along with developing a fast-access technique to allow partial decompression of the target genome in local regions of interest.  From a theoretical viewpoint it will be interesting to understand whether and under what assumptions our algorithm approaches the fundamental limit on compression.

\section*{Acknowledgment}
The authors thank Chris Miller, Itai Sharon and Golan Yona for helpful discussions. This  work was partially supported by the Center for Science of Information (CSoI), an NSF Science and Technology Center, under grant agreement CCF-0939370, a Stanford Graduate Fellowship, NDSEG and La Caixa fellowship program.


\begin{thebibliography}{1}


\bibitem{nature}
E. S. Lander, et al., "Initial sequencing and analysis of the human genome", Nature, vol. 409, pp. 860-921, 2001.
\bibitem{wheeler}
D. A. Wheeler, et al., "The complete genome of an individual by massively parallel DNA sequencing", Nature, vol. 452, pp. 873-876, 2008.
\bibitem{dnazip}
S. Christley, Y. Lu, C. Li, and X. Xie, "Human Genomes as email attachments", Bioinformatics, vol. 25, pp.274-275, 2008.
\bibitem{green}
A. J. Pinho, D. Pratas and S. P. Garcia, "GReEn: a tool for efficient compression of genome resequencing data", Nucleic Acids Research, December 2011.
\bibitem{grs}
C. Wang and D. Zhang, "A novel compression tool for efficient storage of genome resequencing data", Nucleic Acids Research, 39, e45, April 2011.
\bibitem{rlz1}
S. Kuruppu, S. J. Puglisi and J. Zobel, "Relative Lempel-Ziv compression of genomes for large-scale storage and retrieval", Proceedings of the 17th international conference on String processing and information retrieval, SPIRE 2010.
\bibitem{rlz2}
S. Kuruppu, S. J. Puglisi and J. Zobel, "Optimized relative Lempel-Ziv compression of genomes", Proceeding of ACSC 2011.

\bibitem{manipulation}
L. S. Heath, A. Hou, H. Xia and L. Zhang, "A genome compression algorithm supporting manipulation", Proc LSS Comput Syst Bioinform Conf., Vol. 9, pp. 38-49, August, 2010.

\bibitem{tse}
N. Ma, K. Ramchandran and D. Tse, "A Compression Algorithm Using Mis-aligned side information", ITA Workshop, UCSD, 2012.


\bibitem{lz}
J. Ziv and A. Lempel, "A universal algorithm for sequential data compression", IEEE Trans. Information Theory, vol. IT-24, pp.337-343, May 1977.

\bibitem{ctw}
F. M. J. Willems, Y. M. Shtarkov and T. J. Tjalkens, "The Context-Tree Weighting Method: Basic Properties", IEEE Trans. Information Theory, vol. 41, no. 3, pp. 653 - 664, May 1995.

\bibitem{lz2}
P. Subrahmanya and T. Berger, "A sliding window Lempel-Ziv algorithm for differential layer
encoding in progressive transmission", Proc.
1995 IEEE Int. Symp. Inf. Theory, Whistler, BC,
Canada, pp. 266, Jun. 1995.


\bibitem{ctw2}
H. Cai, S. Kulkarni and S. Verd\'{u}, "An Algorithm for Universal Lossless Compression with Side Information", IEEE Trans. Information Theory, vol. 52, no. 9, pp. 4008-4016, Sep. 2006.

\bibitem{huffman}
D. A. Huffman, et al., "A method for the construction of minimum-redundancy codes", Proc. I.R.E., vol. 40, pp. 1098-1102, Sept. 1952.

\bibitem{golomb}
S. W. Golomb, "Run-length encodings", IEEE Trans. Information Theory, vol. 12, no. 3, pp.399-401, Jul. 1966.
 
\bibitem{korean}
S. M. Ahn, T. Kim, S. Lee, et al. "The first
Korean genome sequence and analysis: full genome sequencing
for a socio-ethnic group", Genome Res., 19, 1622-1629, 2009.

\bibitem{chinese}
J. Wang, W. Wang, R. Li, et al., "The diploid genome
sequence of an Asian individual", Nature, 456, 60-66, 2008.

\bibitem{tair}
E. Huala, A. W. Dickerman, M. Garcia-Hernandez, et al., "The Arabidopsis Information Resource
(TAIR): a comprehensive database and web-based information
retrieval, analysis, and visualization system for a model plant", Nucleic Acids Res., 29, 102-105, 2001.

\bibitem{tigr}
S. Ouyang, W. Zhu, et al., "The TIGR Rice Genome Annotation Resource: improvements
and new features", Nucleic Acids Res., 35, D883-D887, 2007.

\end{thebibliography}
\end{document}